\documentclass[sigconf]{acmart}

\usepackage{color}
\usepackage{subfig}
\usepackage{graphicx}
\AtBeginDocument{%
  \providecommand\BibTeX{{%
    \normalfont B\kern-0.5em{\scshape i\kern-0.25em b}\kern-0.8em\TeX}}}

\copyrightyear{2024}
\acmYear{2024}
\setcopyright{acmcopyright}\acmConference[NANOCOM '24]{The Eleventh Annual ACM
International Conference on Nanoscale Computing and Communication (ACM NanoCom 2024)}{October
28--30, 2024}{Milan, Italy}
\acmBooktitle{The Eleventh Annual ACM International Conference on Nanoscale
Computing and Communication (NANOCOM '24), October 28--30, 2024, Milan,
Italy}
\acmPrice{XX.XX}
\acmDOI{XX.XXXX/XXXXXXX.XXXXXX}
\acmISBN{XXXXXXXXXXXX}

\begin{document}

\title[A comparative analysis of preamble sequence for Galvanic Coupling Intra-Body Communication]{A comparative analysis of preamble sequences for Galvanic Coupling Intra-Body Communications}

\author{Farzana Kulsoom}
\affiliation{%
	\institution{University of Engineering and Technology}
	\city{Taxila}
	\country{Pakistan}}
\email{farzana.kulsoom@uettaxila.edu.pk}

\author{Pietro Savazzi}
\affiliation{%
	\institution{University of Pavia\\ $\&$ CNIT Consorzio Nazionale Interuniversitario per le Telecomunicazioni - Unità di Pavia}
	\city{Pavia}
	\country{Italy}}
\email{pietro.savazzi@unipv.it}

\author{Fabio Dell'Acqua}
\affiliation{%
	\institution{University of Pavia\\ $\&$ CNIT Consorzio Nazionale Interuniversitario per le Telecomunicazioni - Unità di Pavia}
	\city{Pavia}
	\country{Italy}}
\email{fabio.dellacqua@unipv.it}

\author{Hassan Nazeer Chaudhry}
\affiliation{%
	\institution{Barani Institute of Information Technology, Rawalpindi}
	\city{Rawalpindi}
	\country{Pakistan}}
\email{hassan@biit.edu.pk}

\author{Anna Vizziello}
\affiliation{%
	\institution{University of Pavia\\ $\&$ CNIT Consorzio Nazionale Interuniversitario per le Telecomunicazioni - Unità di Pavia}
	\city{Pavia}
	\country{Italy}}
\email{anna.vizziello@unipv.it}



\begin{abstract}
  Galvanic coupled-intra-body communication (GC-IBC) is an innovative research area contributing to transform personalized medicine by enabling seamless connectivity and communication among implanted devices. To establish a reliable communication link between implanted devices, the preambles play a crucial role by e.g. conveying syncronization information or supporting channel response estimation. The preambles are carefully designed to ensure that they are mutually orthogonal, to minimize self-interference and maximize separability. For that purpose, many permeable sequences are proposed in the literature for 5G and sensor networks. Golay code, Constant Amplitude Zero Auto Correlation (CAZAC) and Zadoff-Chu (Z-Chu) sequences are among the most popular ones. In this work, we performed a comparative analysis of these sequences to determine their suitability for the GC-IBC system. We evaluated the effectiveness of the preamble sequences on the basis of their correlation properties and probability of error.
\end{abstract}




\maketitle

\section{Introduction}

Next-generation healthcare with real-time physiological data monitoring will be ensured by novel energy-efficient communication techniques capable of interconnecting implanted devices, such as the galvanic coupling (GC) technology \cite{Seyedi2013}.
GC employs a pair of electrodes as transmitter and receiver, which may be attached or implanted in the body, conveying weak ($<$1 mA) electrical currents modulated with data \cite{Swaminathan2017} at low/medium frequencies (1 KHz-100MHz). 

Previous papers on GC analyzed the body channel model mainly using finite-element methods and equivalent  circuit-analysis-based modeling \cite{Seyedi2013}. 
A limited amount of research was spent on defining a channel impulse response (CIR) to characterize the body channel \cite{Tomlinson2015}, which is instead fundamental to the design of a communication system. 

In \cite{Vizziello_channel_2024}, a correlative channel sounding method is used \cite{Papazian2011,Tomlinson2015} employing pseudorandom noise (PN) sequences to evaluate the GC channel impulse response through experimental measurements. While \cite{Tomlinson2015} considered GC frequencies in the range 100 KHz - 1 MHz, \cite{Vizziello_channel_2024} evaluated the band up to 100 KHz, so signals can be transmitted in the baseband, avoiding any modulation scheme and complex synchronization method at the receiver, which is suitable for implanted devices.

To ensure a reliable communication link between implanted devices, the preamble is essential. Preambles are meticulously crafted based on orthogonality principles, guaranteeing that the sequences are mutually orthogonal. Numerous preamble sequences are documented in the literature for 5G and sensor networks. The most popular ones include Golay code \cite{Wysocki2001}, Constant Amplitude Zero Auto-Correlation (CAZAC) \cite{Wen2006}, and Zadoff-Chu (Z-Chu) \cite{Andrews2023} sequences.

In this work, a comparative evaluation of different preamble sequences is taken into account, to determine the best one for use in GC-IBC systems. We evaluated the performance of the preamble sequences on the basis of their correlation characteristics and error probability.

\section{System Model and Architecture}
 \label{sysmodel}
 
 The used system handles all aspects of communication while considering a simulated GC channel. As described in the research work \cite{Vizziello_channel_2024,kulsoom_synch2024} a GC-based system possesses AWGN channel's properties. Therefore, in this work, we have considered the AWGN channel while investigating the impact of preamble sequences on GC-based communication.

\subsection{ Transmitter blocks}
\label{TX}

\begin{table}
	\caption{Parameters setting}\label{tab1}
	\begin{tabular}{|l|c|}
		\hline
		$\mathbf{Parameter}$ & $\mathbf{Value}$\\  
		\hline
		Carrier frequency $f_c$ (kHz) & $\textcolor{black}{10}$\\ 
		Waveform sampling frequency $f_{s}$ (kHz) & $48$ \\ 
		Oversampling frequency $f_s$ (\# of samples) & $\textcolor{black}{16}$\\ 
		Sampling time $T_s$ (ms) & $\textcolor{black}{0.66}$\\
		RX oversampling frequency $f_{s_{rx}}$ (\# of samples) & $2$\\ 
		roll-off of TRX filters $R$ & $0.2$\\ 
		delay of TRX filters  $D$ (\# of samples) & $8$\\
		QAM modulation order $M$ & $2$\\
		RX Wiener filter length $N_f$ (\# of samples) & $11$\\ 
		\textcolor{black}{Modulated sequence $N$ (\# of symbols)} & \textcolor{black}{$10000$}\\
		\textcolor{black}{Preamble length $N_{pre}$ (\# of symbols)} & \textcolor{black}{$1000$}\\
		\hline
	\end{tabular}
\end{table}


This research work uses burst mode transmission; each symbol in the burst comprises the preamble 
followed by the data. 
For simulation, the signal $x$ is generated randomly for each symbol and then the preamble is inserted. A binary phase-shift keying (BPSK) modulation is applied to each symbol. 

We use a low-order modulation, such as BPSK, since intrabody communication is low powered and BSPK is energy efficient, has low complexity, and is low cost \cite{VizzielloGC2020}. Once the data have been modulated using BPSK, the signal is upsampled by the factor of 16, and then pulse shaping is performed using a squared root-raised-cosine (SRRC) filter, as summarized in Table \ref{tab1}. Baseband samples $x(nT_s)$ are produced with a sampling time $T_s$ such that $n=0,1,...,N f_s -1$. 

The transmitted signal is obtained:
\begin{equation}
s(nT_{s})= x(n T_{s}) \cos (2 \pi f_c n T_{s}) 
\end{equation} 
where 
$T_{s}=1/f_{s}$, $f_{s}=48$ KHz and $f_c$ is the carrier frequency.



\subsection{Preamble generation in packets}
The preamble is important in communication systems to detect the presence of packets, coarse signal synchronization, active gain control (AGC) and other communication aspects.

\textcolor{black}{The length of the preamble can affect the system performance. A shorter preamble improves data transmission efficiency, but the design and choice of the preamble must ensure that synchronization and error detection capabilities are not compromised. Optimizing the preamble length is essential for energy-efficient communication systems \cite{Bagci2016}, especially in a battery powered device such as in GC-system.}
Several preamble schemes are used in the state of the art including CAZAC \cite{Wen2006}, Golay codes \cite{Wysocki2001}, and a variant of CAZAC known as Z-Chu \cite{Andrews2023}.  The following subsection discusses the aforementioned preamble schemes.  
  
\subsubsection{Preamble Sequences}

CAZAC sequences are widely considered ideal candidates for preamble generation in communication systems.
A CAZAC sequence $p[l]$ with length $L$ can be represented as:

\begin{equation}
    p[l]=e^j{\frac{2 \pi k l^2}{L}}
\end{equation}

Where $l$ is the sequence index and $k$ is a constant integer. These sequences have specific properties that make them highly effective for preamble detection. For example, their constant amplitude and zero autocorrelation. Each element in a CAZAC sequence has the same magnitude. This means that the power of the signal is evenly distributed, avoiding spikes in power. As a result, the Peak to Average Power Ratio (PAPR) is low. A low PAPR is crucial for maintaining the linearity of power amplifiers. When the power amplifier operates linearly, it reduces distortion and enhances the quality of the transmitted signal. The autocorrelation equation for the CAZAC sequence is as follows:

\begin{equation}
    R_p[m]=
    \begin{cases} 
        \sum^{L-1}_{l=0} p[l] p^*[l-m] & \text{for } m \neq 0 \\
        0 & \text{otherwise}
    \end{cases}
\end{equation}
 
CAZAC sequences are designed to have zero autocorrelation functions for all non-zero shifts. In practical terms, this means that the sequence does not correlate with itself at any offset other than zero. This property is beneficial for synchronization and channel estimation, as it minimizes interference and makes it easier to accurately detect the start of the sequence. \\

\subsubsection{Golay Sequences}
Golay sequences are another type of sequence used in communication systems, known for their unique properties that are beneficial for various applications such as error correction and synchronization. A pair of Golay sequences, $A$ and $B$, have a special correlation property: their autocorrelations sum to zero for all non-zero shifts. Mathematically, if $a[n]$ and $b[n]$ represent a pair of Golay sequences, then the sum of their autocorrelations at any non-zero shift $m$ is zero, as described by the equation below:

\begin{equation}
R_{a}[m] + R_{b}[m] = 0 \quad \text{for } m \neq 0
\end{equation}

Here, $R_{a}[m]$ and $R_{b}[m]$ are the autocorrelation functions of sequences $a[n]$ and $b[n]$, respectively. This property makes Golay sequences extremely useful in scenarios where minimizing interference and ensuring reliable detection are critical. Golay sequences are also constant amplitude sequences, which helps in maintaining a low Peak to PAPR. This is crucial for the efficient operation of power amplifiers in communication systems, similar to CAZAC sequences. In summary, CAZAC sequences feature zero autocorrelation and constant amplitude, making them ideal for synchronization and channel estimation due to low PAPR and minimal interference. Golay sequences come in pairs with complementary autocorrelation, useful for precise synchronization and error correction, also maintaining low PAPR and minimal signal distortion. CAZAC is more complex to generate but excels in high-noise environments, while Golay is simpler to implement and effective in reducing transmission errors. \\

\subsubsection{Z-Chu Sequences}
\textcolor{black}{Z-Chu sequences are another type of sequence widely used in communication systems, known for their unique properties that make them valuable for applications such as synchronization, channel estimation, and random access. These sequences are complex-valued and have a constant amplitude, which ensures a low PAPR. This characteristic is beneficial for the efficient operation of power amplifiers, similar to CAZAC and Golay sequences. Z-Chu sequences are defined mathematically by:}

\begin{equation}
Z_u[l] = \exp\left(-j \frac{\pi u l (l + 1)}{L}\right) \quad \text{for } l = 0, 1, \ldots, L-1 
\end{equation}

\textcolor{black}{where $u$ is a coprime integer with $L$, One of the standout properties of Z-Chu sequences is their zero cyclic autocorrelation for any non-zero shift:}

\begin{equation}
R_{Z_u}[m] = \sum_{l=0}^{L-1} Z_u[l] Z_u^*[l+m] = 0 \quad \text{for } m \neq 0
\end{equation}

\textcolor{black}{This property implies that Z-Chu sequences are perfectly suited for minimizing interference in communication systems, which is critical for reliable detection and synchronization. In addition to their autocorrelation properties, Z-Chu sequences maintain orthogonality among different sequences \cite{Andrews2023}. This orthogonality means that different sequences can be used simultaneously within the same frequency band without interfering with each other, enhancing system capacity and efficiency. Z-Chu sequences are notably complex to generate due to their mathematical formulation, but they excel in high-noise environments and are extensively used in 4G LTE and 5G NR systems for synchronization signals and reference signals. Their robustness in adverse conditions and their ability to maintain low PAPR make them indispensable in modern communication technologies.\\
In summary, Z-Chu sequences, like CAZAC sequences, exhibit zero autocorrelation and constant amplitude, making them ideal for synchronization and channel estimation. They are particularly effective in high-noise environments due to their complex structure and orthogonality, which minimizes interference. While their generation is more challenging compared to Golay sequences, their superior performance in maintaining signal integrity and reducing errors makes them a preferred choice in advanced communication systems.}

\subsection{Receiver blocks for synchronization and signal recovery}

A local oscillator is used for the receiver's baseband signal recovery, matching the frequency and phase of the modulating carrier at the transmitter. Phase errors can degrade transmission performance, and a carrier synchronizer \cite{Mengali1997} is needed as a countermeasure. 
Once the signal is received, a digital down-conversion is carried out by multiplying the received sampled signal $\mathbf{s}_{rx}$ by $\cos (2 \pi f^{'}_{c} n T_{s})$ and $\sin (2 \pi f^{'}_{c} n T_{s})$, where $f^{'}_{c}$ is the carrier frequency at the receiver, thus obtaining $d^i(n T_{s})$ and $d^q(n T_{s})$, respectively.
The sequence $d(n T_{s})=d^i(n T_{s})+jd^q(n T_{s})$ passes through an SRRC filter, resulting in the received baseband signal $r(nT_{s})$ for $n=0,1,...,f_sN-1$. Differences in local oscillator frequencies between the transmitter and receiver , $f_{c}$ and $f^{'}_{c}$, cause carrier synchronization issues. After SRRC, the signal $\mathbf{r}$ is decimated by eight, resulting in two samples per symbol ($f_{s_{rx}}=2$ as shown in Table \ref{tab1}). The resulting sampled signal is:
\begin{equation}
r(nT_s)=\tilde{s}(nT_s-\tau) e^{(2\pi \Delta f n+\theta)}+v(nT_s)
\label{eq_received}
\end{equation}
where $n=0,1,\dots ,f_{s_{rx}}N-1$, $\tilde{s}(nT_s)$ is the envelope samples of the passband signal, $v(nT_s)$ is the sampled AWGN, $\tau$ is the timing offset, $\Delta f=f_c-f^{'}_c$ is the carrier frequency offset (CFO), and $\theta$ is the random phase noise.

In the following we describe the methods used for frequency and timing offset recovery \cite{kulsoom_synch2024}.
A frequency recovery method estimates the frequency offset $\Delta f=f_c-f^{'}_c$. The estimated $\Delta f$ is used to counter-rotate the signal by $2\pi\Delta f$ to tackle frequency synchronization problems:
\begin{equation}
y(nT_s)=r(nT_s) e^{-j2\pi\Delta fnT_s}
\label{eq_controrotation}
\end{equation}

\begin{figure*}[h]
\centering
\includegraphics[width=150mm,height=65mm]{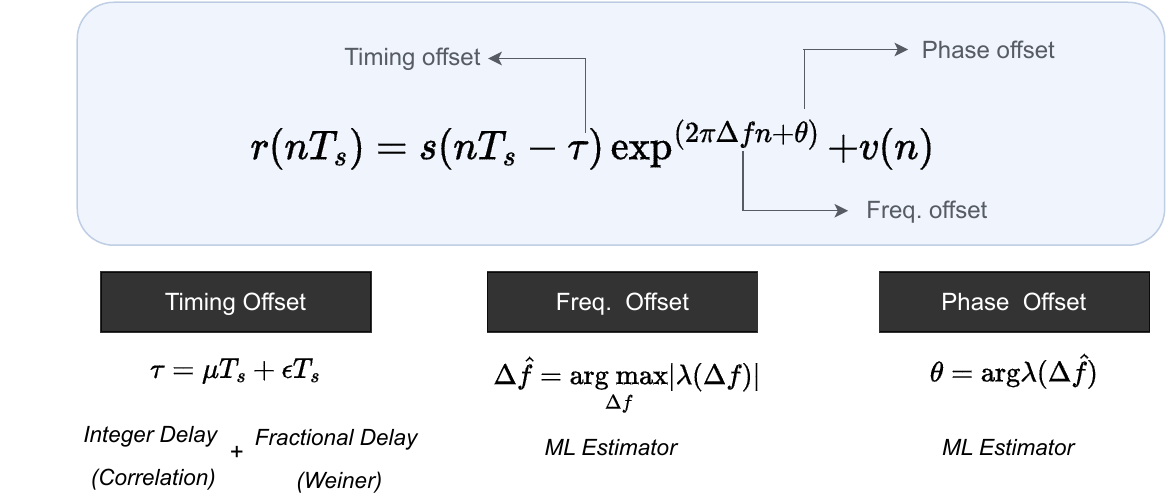}
\caption{\textcolor{black}{Synchronization and offsets in received signals \cite{kulsoom_synch2024}}}
\label{fig:syncproblems}
\end{figure*}

The timing offset $\tau$ in (\ref{eq_received}) is further divided into an integer delay $\mu > 0$ and fractional delay $-0.5 < \epsilon < 0.5$:
\begin{equation}
\tau= \mu T_s+\epsilon T_s
\label{eq_Timg}
\end{equation}
Integer and fractional delays pose limitations in symbol detection. In burst-mode communication, data is transmitted repeatedly after specified intervals to preserve energy. 
To detect the burst start $\hat{\mu}$, a cross-correlation between the received signal and the known preamble is calculated. A Wiener filter is then used for equalization and fractional delay estimation. 

\subsubsection{Step-1: Coarse Carrier Offset Estimation}
Maximum Likelihood (ML) function is used to estimate $\Delta f$ and $\theta$ \cite{Hosseini_2024}:
\begin{equation}
\wedge (r;\tilde{f}, \tilde{\theta)} \approx \Re\big[ e^{-j \tilde{\theta}}\sum_{n=0}^{NL-1} e^{-j(2 \pi n \tilde{f})} r[n] s^*_{\epsilon}[n]\big]
\label{eq_LLF}
\end{equation}
where $L$ is the length of the preamble, $\tilde{f}=\Delta f T_s/N$ is the frequency normalized with respect to the sampling frequency, $s^*_{\epsilon}[n]$ is the sampled version of $s(t-\epsilon T_s) $ at $t=n T_s/N$.
The LF maximization for trial values $\tilde{\theta}$ results in:
\begin{equation}
\tilde{\theta}= \text{arg}\, {\lambda (\tilde{f})}
\end{equation}
\begin{equation}
\hat{ f}= \underset{\tilde{f}}{\text{argmax} }| \lambda (\tilde{f})|
\label{eq:max_freq}
\end{equation}
where $\lambda (\tilde{f})$ represents the likelihood function \cite{kulsoom_synch2024}.  
The phase offset is obtained as:
\begin{equation}
\hat{\theta}= \text{arg} \,{\lambda (\hat{f})}
\end{equation}

\subsubsection{Step-2: Fine CFO Estimation}
After obtaining the coarse estimate, the fine frequency offset is estimated by using interpolation function as follows:
\begin{equation}
\Delta f= \frac{h_{(m-1)}-h_{(m+1)}}{(2h_{(m-1)})-4h_m+2h_{(m+1)}} f_{\text{res}}
\label{quad_inter}
\end{equation}
in which $f_{\text{res}}$ is the frequency resolution obtained by $f_{\text{res}}= \hat{f}_m-\hat{f}_{m-1}$, where $f_m$ is the maximizing frequency calculated from (\ref{eq:max_freq}), and $h_m$ and $h_{m+1}$ represent the energies at these maximizing frequencies $\hat{f}_m$ and $\hat{f}_{m+1}$, respectively. 

\subsubsection{Step-3: Timing Offset Estimation}
The burst start, $\hat{\mu}$, is detected by cross-correlating the counter-rotated received signal $y$ and the preamble $p$:
\begin{equation}
\hat{\mu}= \arg\max_n \left| \sum_{k=0}^{L-1}y((n+k)T_s)p^*(kT_s)\right|
\label{eq_delay}
\end{equation}

The Wiener filter equalizes the signal and also estimates the fine time delay $\hat{\epsilon}$- The Wiener coefficients are calculated by minimizing the mean squared error (MSE) between the transmitted and estimated
preamble:
\begin{equation}
\label{eq_MSE}
e=E\left\{\left[p(nT_s)-\hat{p}(nT_s)\right]^2\right\}
\end{equation}
The vector $\mathbf{w}$ of the Wiener coefficients are calculated through the well-known Wiener-Hopf equation:
\begin{equation}
\label{eq_Wiener}
\mathbf{w}=\mathbf{A}_{y_p}^{-1}\mathbf{a}_{py_p}
\end{equation}
where $\mathbf{A}_{y_p}$ is the autocorrelation matrix of the counter-rotated received preamble $y_p$, and $\mathbf{a}_{py_p}$ is the cross-correlation vector between the known transmitted preamble $p$ and the received one $y_p$ \cite{kulsoom_synch2024}. Once the weights of the Wiener filter are calculated, they are used to estimate the transmitted sequence $\hat{c}_{tx}(nT_s)$:
\begin{equation}
\label{eq_outW}
\hat{c}_{tx}(nT_s)=\sum_{l=0}^{N_f-1}w(lT_s)y((n-l)T_s+\hat{\mu} )
\end{equation}

in which $\hat{\mu}$ is the estimate of the sample index where the reference preamble starts.
After down-sampling to one sample per symbol, the BPSK decision is performed using $\Re(\hat{c}_{tx}(nT_s))$, reconstructing the original signal at the receiver.

\section{Results and Discussion}
This section analyzes the impact of various preamble sequences on system performance and present the findings. For that purpose, three popular preambles CAZAC, Z-Chu, and Golay code have been transmitted with data individually. For all results, BPSK modulation is considered and for estimating synchronization parameters the following methods are used: a Maximum Likelihood (ML) estimator for phase and frequency, a correlation method for coarse timing delay estimation, a Wiener filter for fine delay estimate. The main motivation for this experiment is to identify the most suitable preamble sequence for GC-IBC system. \textcolor{black}{Different preamble lengths, measured in number of samples, are considered.} \\
  \begin{figure*}
\subfloat[\label{fig7:a}][Received signal with Golay code as preamble after applying filter ]
{\includegraphics[width=.5\linewidth,height=4.8cm]{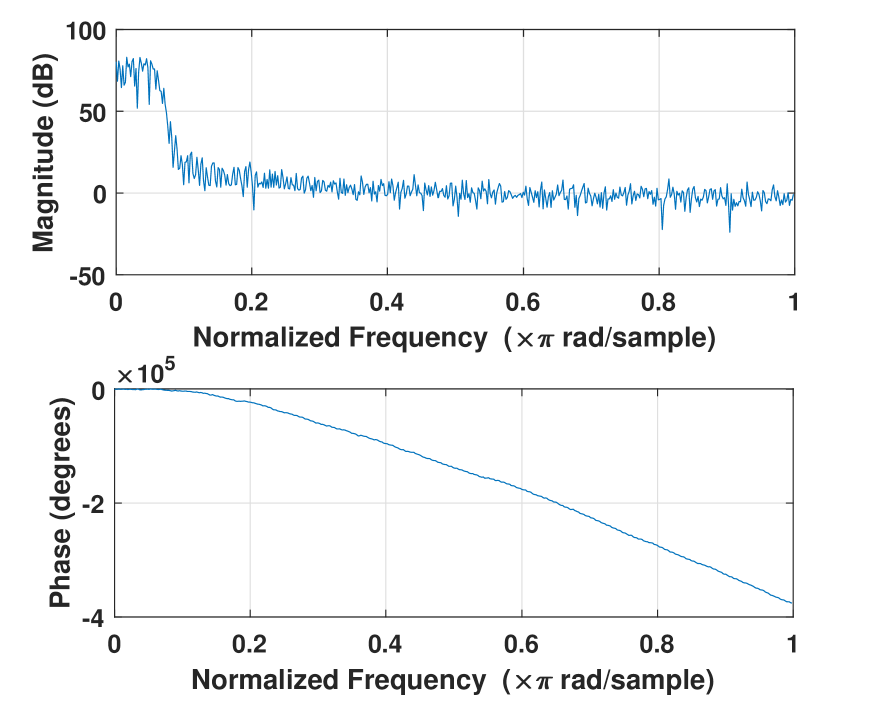}}
\subfloat[\label{fig7:b}][Received signal with Golay code as preamble after CFO Compensation]
{\includegraphics[width=.5\linewidth,height=4.8cm]{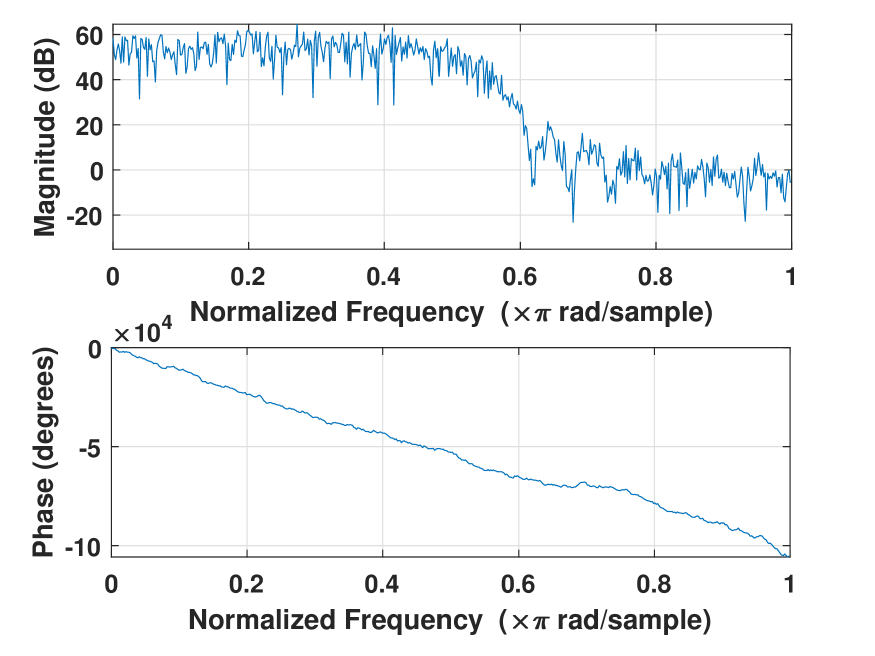}}
 \caption{Received signal with Golay code as preamble at SNR=8dB, preamble length=64}
 \label{fig_Freq_cazac}
 \end{figure*}

\begin{figure*}
\subfloat[\label{fig7:a}][Received signal with CAZAC code as preamble after applying filter ]
{\includegraphics[width=.5\linewidth,height=4.8cm]{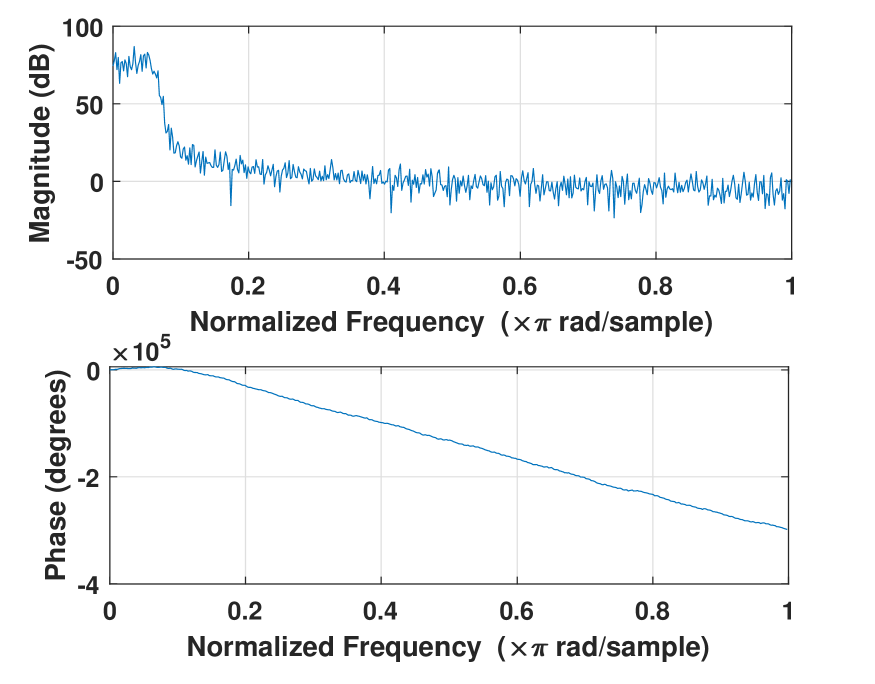}}
\subfloat[\label{fig7:b}][Received signal with CAZAC code as preamble after CFO Compensation]
{\includegraphics[width=.5\linewidth,height=4.8cm]{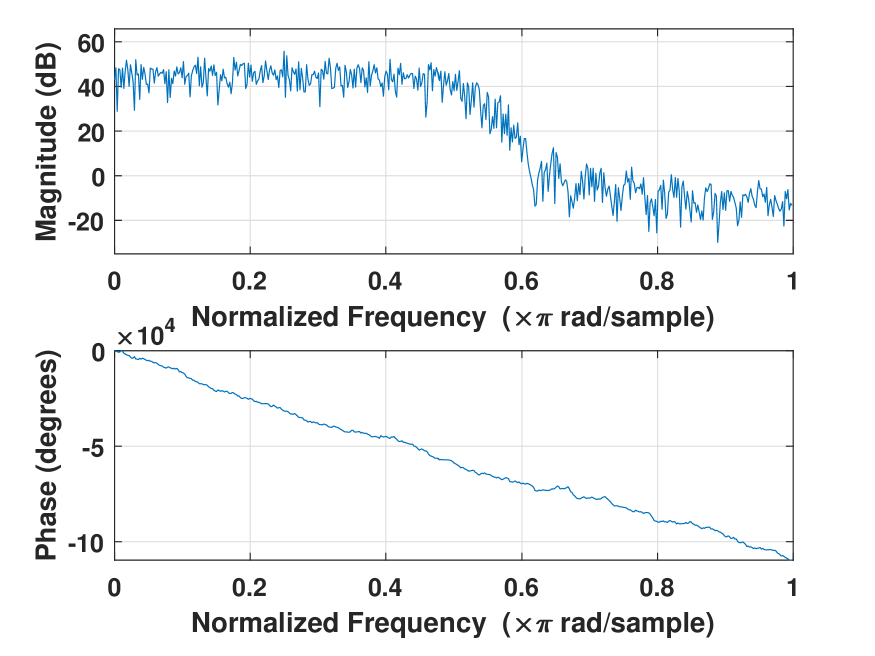}}
 \caption{Received signal with CAZAC as preamble at SNR=8dB, preamble length=64}
 \label{fig_Freq_golay}
 \end{figure*}

Figs. \ref{fig_Freq_cazac} and \ref{fig_Freq_golay} provide a step-by-step frequency-domain analysis for Golay and CAZAC codes. In part (a) of both figures comparison of phase and magnitude between the transmitted and received signals with filtering is shown, and then part (b) of the figures depicts the phase and magnitude responses after CFO compensation. The phase and frequency responses of the Cazac code exhibit slightly more distortions compared to the Golay code. 

\begin{figure}[htbp]
	\centering
{\includegraphics[width=0.5\textwidth]{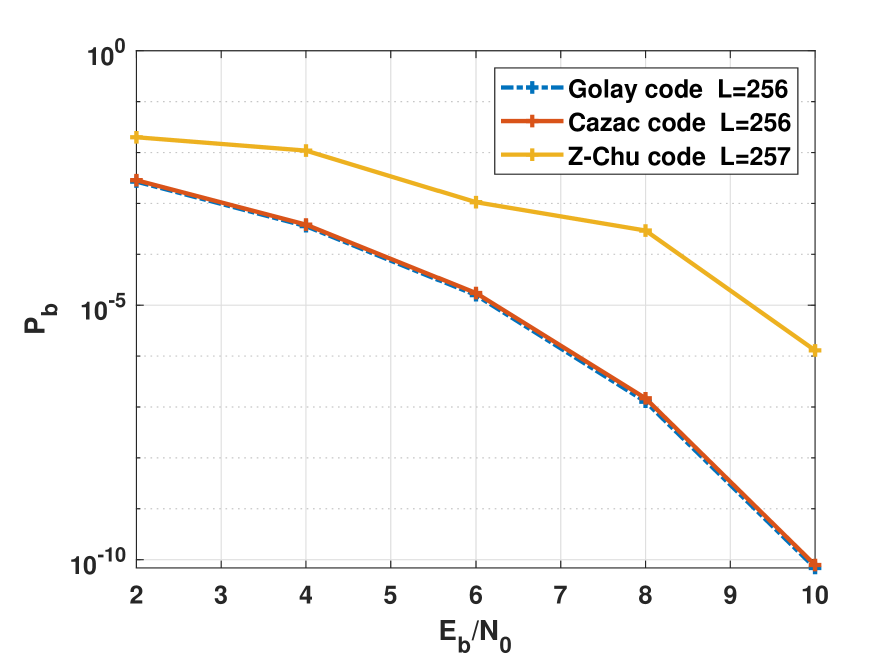}}
	\caption{Probability of error VS $E_b$/$N_0$ for different preamble sequences}
\label{Fig:compar_all_pre}
\end{figure}

\begin{figure}
	\centering	{\includegraphics[width=0.5\textwidth]{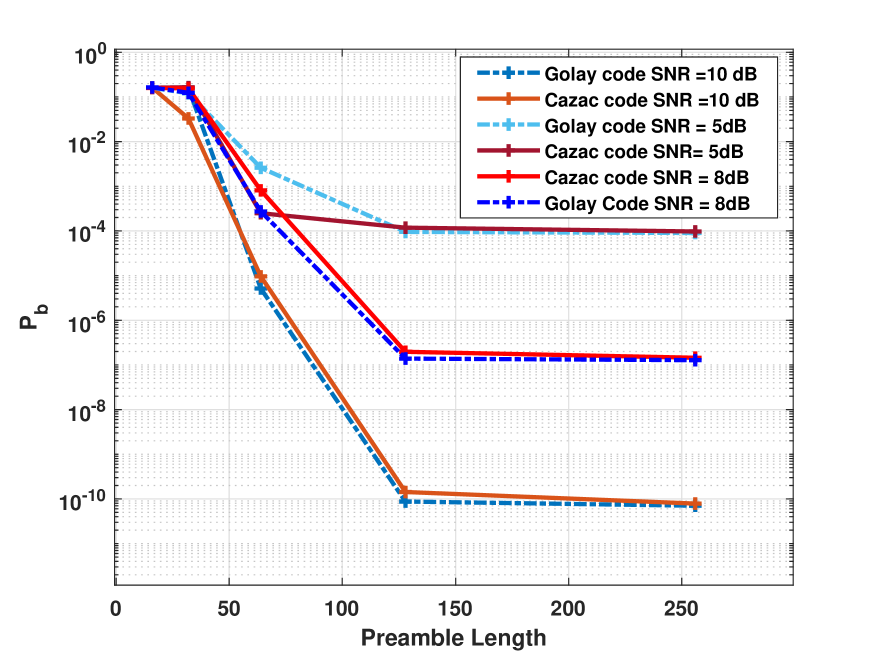}}
	\caption{Comparison of Preamble sequences for varying $E_b$/$N_0$}
    \label{Fig:compar_length}
\end{figure}

In Figs. \ref{Fig:compar_all_pre} and \ref{Fig:compar_length} a comparison of the probability of error $P_b$ is presented for the three aforementioned preamble codes while varying $E_b/N_0$. The probability of error has been calculated by using the following;

\begin{equation}
P_b=E\left(\left\|\frac{\mathbf{c}_{tx_{data}}}{\sqrt{P_{tx}}}-\frac{\Re(\hat{\mathbf{c}}_{tx_{data}})}{\sqrt{P_{rx}}}\right\|^2\right),  \phantom{xxxxxi} 
\end{equation}

in which $\mathbf{c}_{tx_{data}}$ is the transmitted sequence after the BPSK mapper, $\hat{\mathbf{c}}_{tx_{data}}$ is the estimated sequence, i. e., the received sequence before the BPSK decision, downsampled at one bit per symbol.  
The sequences at the transmitter and receiver are normalized to the power $P_{tx}$ and $P_{rx}$, where $P_{tx}=\sum_{n=1}^{N}(c_{tx_{data}}(n))^2/N$ and $P_{rx}$ is equal to $\sum_{n=1}^{N}(\Re(\hat{c}_{tx_{data}}(n)))^2/N$.

In Fig. \ref{Fig:compar_all_pre}, CAZAC, Z-Chu and Golay codes are shown. Although Z-Chu sequence has perfect autocorrelation properties and low PAPR, the performance of Z-Chu sequences in various applications can be constrained by their limited lengths, which are restricted to specific values such as prime numbers. In Fig. \ref{Fig:compar_all_pre}, the performance of the Z-Chu sequence is worse compared to Golay and CAZAC. On the other hand, for a sequence length of 256 \textcolor{black}{number of samples}, the performance of both codes is almost the same, with a slight edge to the Golay code. \\

In Fig. \ref{Fig:compar_length}, comparisons of preamble sequences are shown with increasing preamble length at a constant SNR ratio.
Since the Z-Chu sequence $P_b$ exhibits high values, as depicted in Fig. \ref{Fig:compar_all_pre}, only Golay and CAZAC sequences have been taken into account in Fig. \ref{Fig:compar_length}. As anticipated, the error rate is higher for a shorter preamble length compared to a longer one. In most scenarios, the Golay code demonstrates superior performance compared to other codes. However, at low signal-to-noise ratios, such as 5dB, the CAZAC sequence surpasses the Golay code specifically for shorter preambles. The main reason is CAZAC sequence's constant amplitude characteristics make it particularly well-suited for high-noise scenarios.
However, since a high SNR value may be obtained in GC channels \cite{VizzielloGC2020}, Golay seems to be the preferable choice.

\begin{figure}
	\centering	{\includegraphics[width=0.5\textwidth]{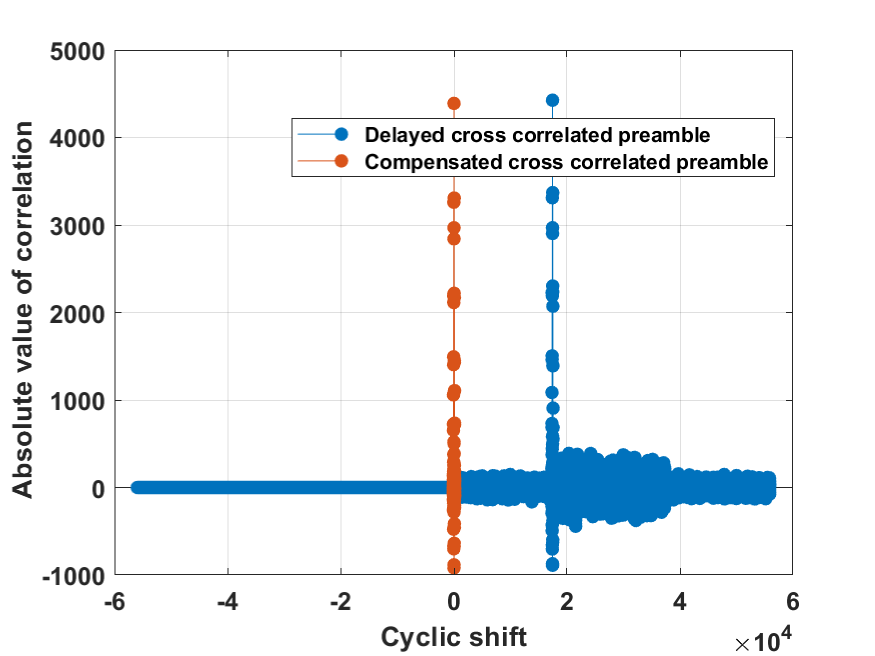}}
	\caption{Cross correlation of transmitted and received preamble for Golay code SNR=8dB, preamble length=64}
 \label{Fig:corr_golay}
\end{figure}

\begin{figure}
	\centering	{\includegraphics[width=0.5\textwidth]{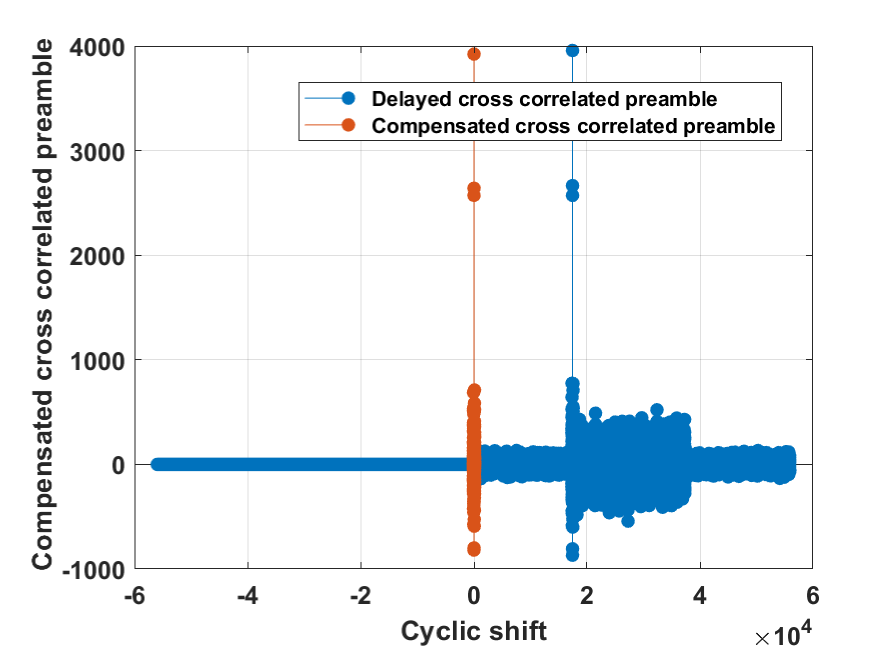}}
	\caption{Cross correlation of transmitted and received preamble for CAZAC code SNR=8dB, preamble length=64}
 \label{Fig:corr_cazc}
\end{figure}
In Fig. \ref{Fig:corr_golay} and \ref{Fig:corr_cazc} the cross-correlation of Golay and CAZAC sequences has been shown for delayed sequences. The Wiener filter and correlation function estimated this delay. Then the cross-correlation of the received delay-compensated signal and transmitted preamble is calculated. The figure shows that The Golay code has a slightly higher peak absolute value of correlation than CAZAC, which indicates a strong similarity.


\section{Conclusion}

In this paper, a comparative analysis of different preamble sequences was conducted to determine their suitability for GC-IBC systems.
Golay code, CAZAC and Z-Chu sequences were investigated and evaluated on the basis of their correlation properties and probability of error. 
Both CAZAC and Golay sequences show very good performance with slightly better results for Golay in most scenarios, while CAZAC performs better in high-noise scenarios. Since the GC channel exhibits a high SNR value \cite{VizzielloGC2020}, Golay is the preferable choice because it is simpler to implement than CAZAC. 
Future research efforts will be devoted to identifying the best modulation and coding schemes for these specific applications, considering that they require energy-efficient solutions. 
Other research directions include implementation of systems to increase the data rate in GC communications
while keeping linear modulation schemes \cite{Vizziello2016,KulsoomReduced2018},
integrated with the development of compressive sensing methods to reduce the complexity of recovering data \cite{Alesii2015}. Specific development of modulation and coding schemes considering multiple-input and multiple-output (MIMO) configurations \cite{Kulsoom2018} are planned to be tailored for intra-body networks. Furthermore, besides a single link establishment, multiple implants scenario will be considered, where it is fundamental
to use opportunistic wake-up methods and localization of the
devices \cite{Kianoush2016,Vizziello2013,Stelzner2019}, which is essential in intra-body networks for
medical applications.

\section*{Acknowledgements}
This publication is part of the project NODES which has received funding from the MUR - European Union – M4C2 1.5 of PNRR funded by the European Union - NextGenerationEU (Grant agreement no. ECS00000036)

\bibliographystyle{ACM-Reference-Format}    


\end{document}